\documentclass[prl,aps,twocolumn,superscriptaddress,nofootinbib,longbibliography]{revtex4-2}
\usepackage{graphicx} 
\usepackage{dcolumn}  
\usepackage{bm}       
\usepackage{amsmath}
\usepackage{epsfig}
\usepackage{color}

\DeclareMathAlphabet\mathbfcal{OMS}{cmsy}{b}{n}

\begin{document}
\title{Green Tensor Analysis of Lattice Resonances in Periodic Arrays of Nanoparticles}

\author{Lauren Zundel}
\affiliation{Department of Physics and Astronomy, University of New Mexico, Albuquerque, New Mexico 87106, United States}
\author{Alvaro Cuartero-Gonz\'alez}
\affiliation{Departamento de F\'{i}sica Te\'orica de la Materia Condensada and Condensed Matter Physics Center (IFIMAC), Universidad Aut\'onoma de Madrid, E-28049 Madrid, Spain}
\author{Stephen Sanders}
\affiliation{Department of Physics and Astronomy, University of New Mexico, Albuquerque, New Mexico 87106, United States}
\author{Antonio  I. Fern\'andez-Dom\'{i}nguez}
\affiliation{Departamento de F\'{i}sica Te\'orica de la Materia Condensada and Condensed Matter Physics Center (IFIMAC), Universidad Aut\'onoma de Madrid, E-28049 Madrid, Spain}
\author{Alejandro Manjavacas}
\email[Corresponding author: ]{a.manjavacas@csic.es}
\affiliation{Instituto de \'Optica (IO-CSIC), Consejo Superior de Investigaciones Cient\'ificas, 28006 Madrid, Spain}
\affiliation{Department of Physics and Astronomy, University of New Mexico, Albuquerque, New Mexico 87106, United States}

\date{\today}

\begin{abstract}
When arranged in a periodic geometry, arrays of metallic nanostructures are capable of supporting collective modes known as lattice resonances. 
These modes, which originate from the coherent multiple scattering between the elements of the array, give rise to very strong and spectrally narrow optical responses. Here, we show that, thanks to their collective nature, the lattice resonances of a periodic array of metallic nanoparticles can mediate an efficient long-range coupling between dipole emitters placed near the array. Specifically, using a coupled dipole approach, we calculate the Green tensor of the array connecting two points and analyze its spectral and spatial characteristics. This quantity represents the electromagnetic field produced by the array at a given position when excited by a unit dipole emitter located at another one.  We find that, when a lattice resonance is excited, the Green tensor is significantly larger and decays more slowly with distance than the Green tensor of vacuum. Therefore, in addition to advancing the fundamental understanding of lattice resonances, our results show that periodic arrays of nanostructures are capable of enhancing the long-range coupling between collections of dipole emitters, which makes them a promising platform for applications such as nanoscale energy transfer and quantum information processing.
\end{abstract}

\maketitle


Lattice resonances are collective modes supported by periodic arrays of nanostructures that originate from the coherent multiple scattering between the individual array constituents \cite{ZKS03,ZJS04,paper090,AB08,KSG08,CSY08,VGG09,VAP14,HMS16,HB16,WRV18,KKB18,CBW19,JTH19,UZR21}. These resonances appear in the spectrum at wavelengths commensurate with the periodicity of the array and, due to their collective nature, produce optical responses that are simultaneously very strong and spectrally narrow, thus leading to record quality factors for metallic systems \cite{AYW10,ZRG17,KLW17,ama59,LZG19,ZIG20,DLP20,ama72,BRM21,ama73}.  Thanks to these exceptional properties, periodic arrays of metallic nanostructures are being used in a broad range of applications. These include the implementation of ultrasensitive sensors \cite{AYA09,TKS14,DTW18,MHG18}, the development of platforms for exploring new physical phenomena \cite{VMR14,TEP16,RFV13,RHF17,HMV18_2}, as well as the design of different optical elements, such as light-emitting devices \cite{RLV12,LLR13,ZDS13,LGV14,SK14_2,RLV16,ZHG16,COK16,SVH17,WYW17,GNH19,VKK19}, lenses \cite{HWB19}, color filters \cite{ama41,KYB17,ETB19}, and nonlinear devices \cite{CKL16,MKA17,HRB18,HKW19}.

\begin{figure*}
\begin{center}
\includegraphics[width=150mm,angle=0]{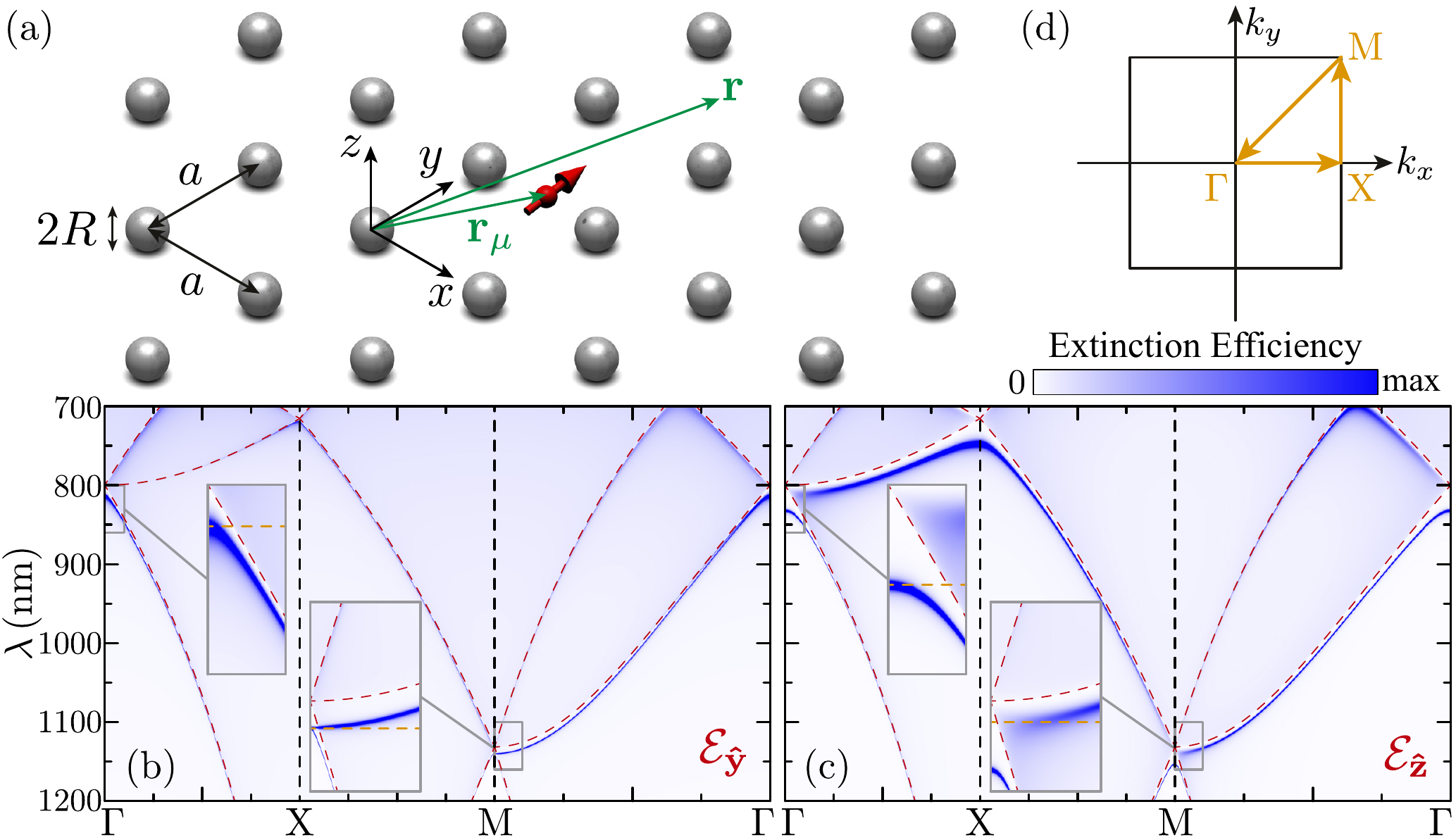}
\caption{ (a)  Schematics of the system under study, consisting of a square array of period $a$ made of identical silver nanospheres with radius $R$. The array is located in the $xy$-plane and is surrounded by vacuum. We are interested in the analysis of the Green tensor of the array connecting two points $\mathbf{r}_{\mu}$ and $\mathbf{r}$, which represents the field produced by the array at $\mathbf{r}=(x,y,z)$ when excited by a unit dipole located at $\mathbf{r}_{\mu}=(x_{\mu},y_{\mu},z_{\mu})$. (b,c) Dispersion diagrams showing the extinction efficiency of an array with $a=800\,$nm and $R=100\,$nm for $y$ (b) and $z$ (c) polarization (see Eq.~\ref{eqExt}), calculated along the path in the first Brillouin zone depicted in (d). The red dashed curves in panels (b) and (c) indicate the position of the Rayleigh anomalies, while the insets show zooms of relevant regions of the dispersion relations, with the yellow dashed lines marking the onset and cutoff of the lowest-order lattice resonance at the $\Gamma$ and M points, respectively.} \label{fig1}
\end{center}
\end{figure*}

Interestingly, in the majority of these applications, the lattice resonances are excited from the far-field using a propagating electromagnetic wave, such as a collimated laser. As a consequence, most of the theoretical characterization of the optical properties of periodic arrays of metallic nanostructures has been focused on describing their far-field response through the analysis of quantities such as the reflectance, transmittance, and absorbance \cite{WRV18,KKB18}. However, lattice resonances also produce very large electromagnetic fields around the array \cite{ZS05,NKD12,HMH16,GVK16}, which, as we have recently shown, are ultimately limited by the number of elements of the array that interact coherently \cite{ama68}. 
The strong near-fields provided by lattice resonances play a crucial role for applications, such as nanolasing, in which the arrays interact with quantum emitters placed in their vicinity \cite{RAM11,ZDS13,WWK18}.  Specifically, in these systems, the lattice resonances couple with the emitters (usually quantum dots or dye molecules) that constitute the gain medium and provide the necessary feedback to achieve lasing \cite{YHD15,HRV17,WYW17,GNH19,FWB19,PHN19,GK19,TMR20,GSL20}. These modes can also strongly influence the emission patterns of the emitters \cite{LG17,HPC19}. Furthermore, the collective character of lattice resonances and their extended nature makes them ideal candidates to provide an efficient long-range interaction between emitters placed near the array. This possibility has started to be explored to achieve collective emission \cite{YBC20,YLL21}, as well as long-range energy propagation \cite{YOW20}.

In this article, motivated by the recent experimental advances, we provide a detailed theoretical investigation of the coupling between dipole emitters mediated by the lattice resonances supported by periodic arrays of metallic nanoparticles. To that end, we implement a theoretical approach based on the coupled dipole model that allows us to compute the Green tensor of the array connecting two points $\mathbf{r}_{\mu}$ and $\mathbf{r}$. This quantity, which represents the electromagnetic field produced by the array at $\mathbf{r}$ when excited by a unit dipole located at $\mathbf{r}_{\mu}$, completely describes the optical response of the array. Using this approach, we analyze the spectral and spatial characteristics of the Green tensor of the array and show that, when a lattice resonance is excited, this quantity is largely enhanced with respect to the Green tensor of vacuum and decays with $|\mathbf{r}-\mathbf{r}_{\mu}|$ at a much slower rate.  Our results contribute to the fundamental knowledge of lattice resonances by providing a full characterization of their interaction with dipole emitters, thus facilitating their use for the enhancement of the long-range coupling between dipole emitters.

\section{Results and Discussion}

The system under study, which is depicted in Figure~\ref{fig1}(a), consists of a square array with period $a$ made of identical silver nanospheres of radius $R$. The array is located in the $xy$-plane and surrounded by vacuum. 
We assume that $R$ is significantly smaller than both $a$ and the wavelength of light $\lambda$, which allows us to characterize the response of the array using a coupled dipole model \cite{ZKS03,paper090,TD12,LK16,ama59,KK19}. Within this approximation, we model each of the nanoparticles of the array as a point dipole with both electric $\mathbf{p}$ and magnetic $\mathbf{m}$ components, whose responses are characterized by an electric $\boldsymbol{\alpha}^{\rm E}$ and a magnetic $\boldsymbol{\alpha}^{\rm M}$ polarizability, respectively.  As shown explicitly in the Appendix, the coupled dipole model allows us to derive the following closed expression for the electric and magnetic field produced by the array at a point $\mathbf{r}=(x,y,z)$, when excited by a unit dipole with electric $\boldsymbol{\hat{\mu}}^{\rm E}$ and magnetic $\boldsymbol{\hat{\mu}}^{\rm M}$ components, located at $\mathbf{r}_{\mu}=(x_{\mu},y_{\mu},z_{\mu})$
\begin{equation}
\left[\begin{matrix}\mathbf{E}(\mathbf{r})\\ \mathbf{B}(\mathbf{r})\end{matrix}\right]
= \frac{a^2}{4\pi^2} \int_{\rm 1BZ} d\mathbf{k}_{\parallel}   \mathbfcal{G}(\mathbf{k}_{\parallel},\mathbf{r})  \mathbfcal{A}(\mathbf{k}_{\parallel})   \mathbfcal{G}(\mathbf{k}_{\parallel},-\mathbf{r}_{\mu})  \left[\begin{matrix} \boldsymbol{\hat{\mu}}^{\rm E} \\ \boldsymbol{\hat{\mu}}^{\rm M} \end{matrix}\right]. \label{eqE}
\end{equation}
Here, ${\rm 1BZ}$ stands for the first Brillouin zone, $\mathcal{G}(\mathbf{k}_{\parallel},\mathbf{r})$ is the lattice sum tensor 
\begin{equation}
\mathbfcal{G}(\mathbf{k}_{\parallel},\mathbf{r}) = \left[ \begin{matrix} \mathbfcal{G}^{\rm EE}(\mathbf{k}_{\parallel},\mathbf{r}) & \mathbfcal{G}^{\rm EM}(\mathbf{k}_{\parallel},\mathbf{r}) \\ -\mathbfcal{G}^{\rm EM}(\mathbf{k}_{\parallel},\mathbf{r}) & \mathbfcal{G}^{\rm EE}(\mathbf{k}_{\parallel},\mathbf{r}) \end{matrix}\right], \nonumber
\end{equation}
with its electric-electric and electric-magnetic components explicitly defined in the Appendix, and $\mathbfcal{A}(\mathbf{k}_{\parallel})=[\boldsymbol{\alpha}^{-1}-\mathbfcal{G}(\mathbf{k}_{\parallel},0)]^{-1}$ is the polarizability of the array (see Eq.~\ref{eqA2}).  This last quantity encodes the intrinsic response of the array, which is determined by the interplay between the response of the nanoparticles, described by the polarizability tensor $\boldsymbol{\alpha}$, and the geometry of the lattice, contained in the lattice sum tensor $\mathbfcal{G}(\mathbf{k}_{\parallel},0)$. 

The integral over $\mathbf{k}_{\parallel}$ in Eq.~\ref{eqE} is the result of the  localized character of the dipole source, which breaks the periodicity of the problem. 
Importantly, the integrand displays narrow features, associated with the lattice resonances of the array, that make it necessary to use an adaptive integration algorithm \cite{hcubature} to perform the integral. 
Furthermore, the lack of periodicity also means that a full numerical solution of Maxwell's equations, using, for instance, a finite element (FEM) or a finite-difference time-domain (FDTD) method, requires performing a similar integration, thus making such computation very challenging. 

We begin our analysis by characterizing the response of the array through the calculation of its extinction efficiency, defined as \cite{ama63,CBW19}
\begin{equation}
\mathcal{E}_{\mathbf{\hat{n}}} = \frac{4\pi k}{a^2} {\rm Im} \{ \mathbf{\hat{n}}\mathbfcal{A}^{\rm EE}(\mathbf{k}_{\parallel}) \mathbf{\hat{n}}^{\dagger} \},\label{eqExt}
\end{equation}
for polarization along $\mathbf{\hat{n}}$. Here, $\mathbfcal{A}^{\rm EE}(\mathbf{k}_{\parallel})$ is the electric-electric component of the polarizability of the array, which is the term that dominates the response of arrays of metallic nanoparticles, such as those analyzed here. Figures~\ref{fig1}(b) and (c) show, respectively, $\mathcal{E}_{\mathbf{\hat{y}}}$ and $\mathcal{E}_{\mathbf{\hat{z}}}$ for an array with $a=800\,$nm and $R=100\,$nm, calculated along the path in the first Brillouin zone depicted in (d). Here and in the remainder of this work, we compute the electric and magnetic components of the polarizability of the nanoparticles from the corresponding dipolar Mie scattering coefficients \cite{paper024} with a dielectric function described using a Drude model $\varepsilon(\omega)=\varepsilon_{\infty}-\omega_{\rm p}^2/(\omega^2+i\gamma\omega)$ with $\varepsilon_{\infty}=5$, $\hbar\omega_{\rm p}=8.9\,$eV, and $\hbar\gamma=37\,$meV \cite{YDS15}. Examining the results shown in Figures~\ref{fig1}(b) and (c), we observe that the array supports different lattice resonances characterized by large values of the extinction efficiency. The lattice resonances appear at slightly larger wavelengths than the Rayleigh anomalies (indicated by the red dashed lines), at which the real part of the lattice sums diverge. In this work, we focus on the lowest-order lattice resonance, which has its onset at the $\Gamma$ point at a wavelength slightly larger than the array period. Furthermore, it displays a cutoff at the M point at a wavelength slightly larger than $\sqrt{2} a$. These two limits, which are indicated in the insets of Figures~\ref{fig1}(b) and (c) with yellow dashed lines, play an important role in the behavior of the array as we discuss later.

\begin{figure}
\begin{center}
\includegraphics[width=75mm,angle=0]{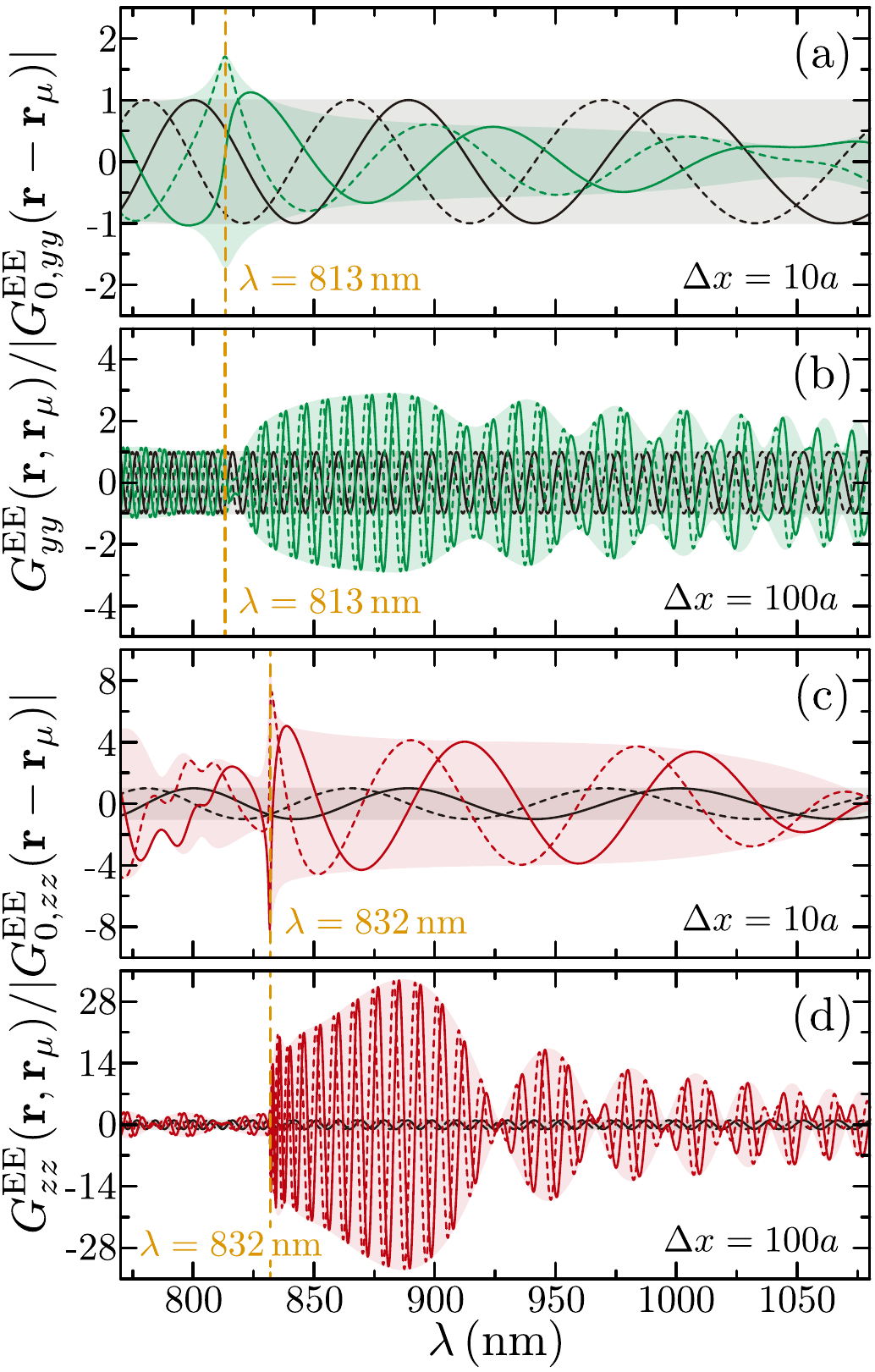}
\caption{Green tensor of a periodic array with $a=800\,$nm and $R=100\,$nm connecting the points $\mathbf{r}_{\mu}=(0,0,2R)$ and $\mathbf{r}=(\Delta x,0,2R)$. 
Panels (a) and (b) show the spectrum for the $yy$ component, while panels (c) and (d) display those for the $zz$ component. In both cases, we analyze the results for $\Delta x = 10a$ (a,c) and $100a$ (b,d). The colored solid and dashed curves represent, respectively,  the real and imaginary parts of the Green tensor, while the shaded areas indicate its envelope.  All of the results are normalized to the amplitude of the same component of the Green tensor of vacuum  $\mathbf{G}^{\rm EE}_{0}(\mathbf{r}- \mathbf{r}_{\mu})$, whose real and imaginary parts are displayed by the black solid curves and whose envelope is signaled by the black shaded areas. In all of the panels, the yellow dashed lines mark the onset of the lattice resonance at the $\Gamma$ point shown in the insets of Figures~\ref{fig1}(b) and (c). } \label{fig2}
\end{center}
\end{figure}

Equation~\ref{eqE} defines the Green tensor of the array connecting two points $\mathbf{r}_{\mu}$ and $\mathbf{r}$ as 
\begin{equation}
\mathbf{G}(\mathbf{r},\mathbf{r}_{\mu})=\frac{a^2}{4\pi^2}  \int_{\rm 1BZ} d\mathbf{k}_{\parallel}   \mathbfcal{G}(\mathbf{k}_{\parallel},\mathbf{r})  \mathbfcal{A}(\mathbf{k}_{\parallel})   \mathbfcal{G}(\mathbf{k}_{\parallel},-\mathbf{r}_{\mu}).\label{eqG}
\end{equation}
Due to its dominant role, in this work, we focus our analysis on the electric-electric component of the Green tensor, which satisfies $\mathbf{E}(\mathbf{r}) = \mathbf{G}^{\rm EE}(\mathbf{r},\mathbf{r}_{\mu}) \boldsymbol{\hat{\mu}}^{\rm E}$ or, in other words, represents the electric field produced by the array at $\mathbf{r}$ when excited by a unit electric dipole placed at $\mathbf{r}_{\mu}$.  
In Figure~\ref{fig2}, we plot the spectrum of the $yy$ (a,b) and $zz$ components (c,d) of $\mathbf{G}^{\rm EE}(\mathbf{r},\mathbf{r}_\mu)$. We assume $\mathbf{r}_{\mu}=(0,0,2R)$ and $\mathbf{r}=(\Delta x, 0, 2R)$ with $\Delta x=10a$ (a,c) and $100a$ (b,d) and the same values for the period of the array and the size of the nanoparticles as in Figure~\ref{fig1}, \textit{i.e.}, $a=800\,$nm and $R=100\,$nm. We use colored solid and dashed curves to represent, respectively, the real and imaginary parts of the Green tensor and a color shaded area for its envelope. The results are normalized to the amplitude of the same component of the Green tensor of vacuum $\mathbf{G}^{\rm EE}_0(\mathbf{r}-\mathbf{r}_{\mu})$ connecting the same two points, which is defined in the Appendix. Notice that this quantity represents the electric field produced at $\mathbf{r}$ by a unit electric dipole placed at $\mathbf{r}_{\mu}$ in absence of the array. For this particular configuration, we have $G^{\rm EE}_{0,yy}(\mathbf{r}-\mathbf{r}_{\mu}) = G^{\rm EE}_{0,zz}(\mathbf{r}-\mathbf{r}_{\mu}) = \exp(ik\Delta x) [k^2(\Delta x)^2 + i k \Delta x -1]/(\Delta x)^3$, with $k=2\pi/\lambda$.
We plot the real and imaginary parts of this quantity using black solid and dashed curves, as well as its envelope with a black shaded area. 
Analyzing the results of Figure~\ref{fig2}, we notice that, in all cases, the real and imaginary parts of the Green tensor of the array display a fast oscillation similar to that of the Green tensor of vacuum, which arises from the factor $\exp(i k \Delta x)$, and therefore is a clear signature of their far-field character. Expectedly, this oscillation is not present in the envelope of the Green tensors. However, the Green tensor of the array stills displays a second, slower, oscillation whose origin is more complex, as we explain later. 

Another important characteristic of the results shown in Figure~\ref{fig2} is the significant change of the Green tensor of the array after the onset of the lattice resonance at the $\Gamma$ point, which is indicated by the yellow dashed lines and coincides with those displayed in the insets of Figures~\ref{fig1}(b) and (c). Clearly, the contribution of the lattice resonance produces a very large enhancement of the amplitude of  $G^{\rm EE}_{zz}(\mathbf{r},\mathbf{r}_{\mu})$, which grows as $\Delta x$ increases. This enhancement is not as pronounced in the case of the $yy$ component; indeed, for $\Delta x= 10a$, the Green tensor of the array is smaller than that of vacuum. The large difference between the $yy$ and $zz$ components is a direct consequence of the nature of the lattice resonances excited in each of these cases. For the $yy$ component, the corresponding lattice resonance has an in-plane character in which the dipoles induced in the nanoparticles oscillate parallel to the array plane and therefore radiate very efficiently in the perpendicular direction. On the contrary, the lattice resonance corresponding to the $zz$ component is an out-of-plane mode in which the dipoles oscillate perpendicular to the plane of the array. Consequently, they only radiate efficiently along the plane of the array, thus minimizing the radiative losses and producing the much larger values of the Green tensor shown in Figure~\ref{fig2}. Indeed, out-of-plane lattice resonances have been investigated in the past for their large quality factors arising from the reduced radiative losses \cite{ZO11,ZHH12,HDT16}. 

In all of the calculations shown in Figure~\ref{fig2}, $\mathbf{r}_{\mu}$ and $\mathbf{r}$ are separated along the $x$-axis. However, in Figure~S1 of the Appendix, we analyze the amplitude of the Green tensor of the array for a similar displacement between $\mathbf{r}_{\mu}$ and $\mathbf{r}$, but, in this case, along the $y$-axis. As expected from the symmetry of the problem, $G^{\rm EE}_{zz}(\mathbf{r},\mathbf{r}_{\mu})$ remains completely unchanged. However, the value for $G^{\rm EE}_{yy}(\mathbf{r},\mathbf{r}_{\mu})$ becomes significantly smaller. The reason is that, in such a case, there is no lattice resonance involved, since they originate from the far-field coupling between the elements of the array, which vanishes along the direction parallel to the dipole moment induced in the nanoparticles.

\begin{figure*}
\begin{center}
\includegraphics[width=150mm,angle=0]{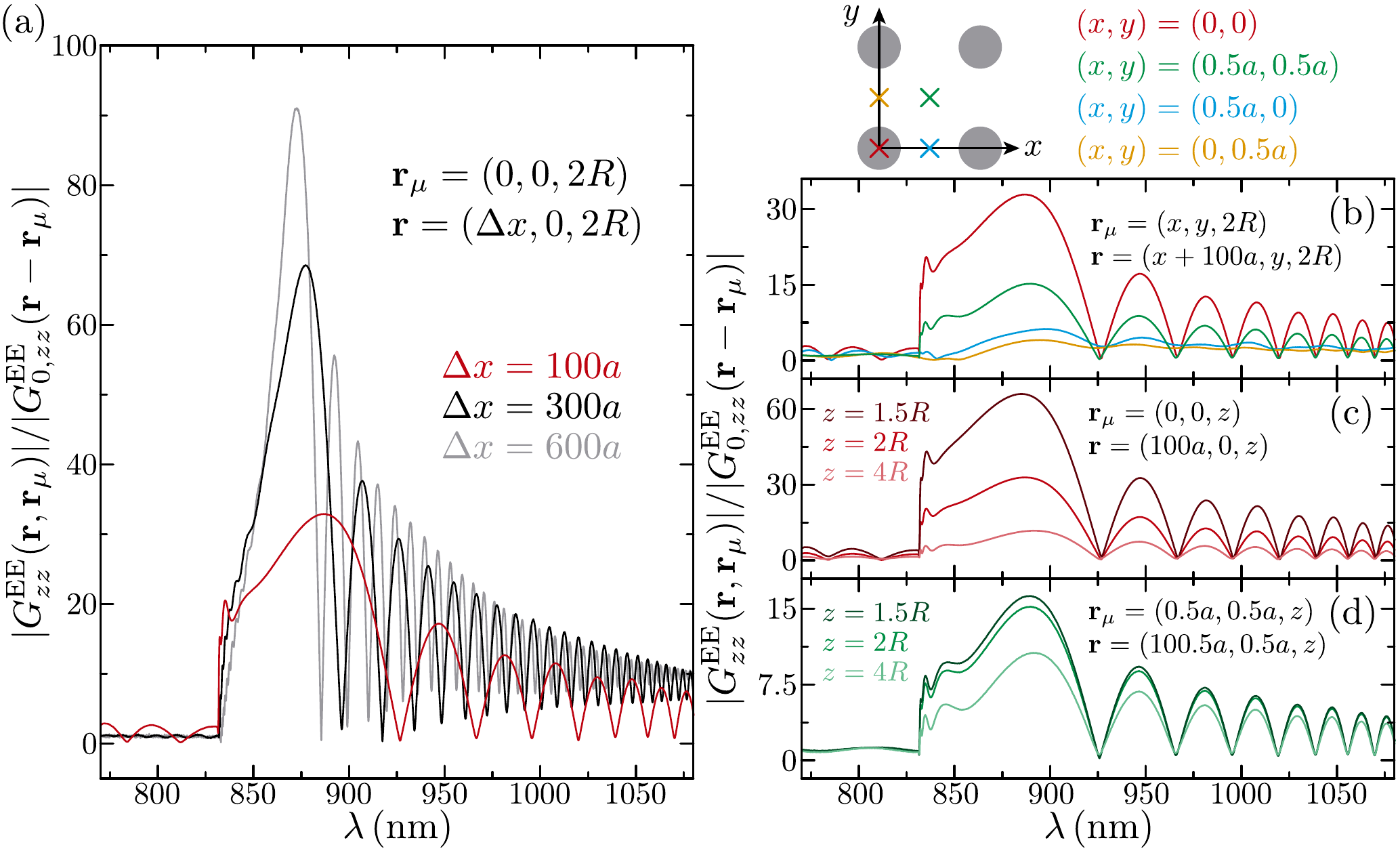}
\caption{Amplitude of the $zz$ component of the Green tensor of a periodic array with $a=800\,$nm and $R=100\,$nm connecting the points $\mathbf{r}_{\mu}$ and $\mathbf{r}$. Panel (a) shows the spectrum for $\mathbf{r}_{\mu}=(0,0,2R)$ and $\mathbf{r}=(\Delta x, 0, 2R)$ with $\Delta x=100a$ (red curve), $300a$ (black curve), and $600a$ (gray curve). Panel (b) shows the spectrum for $\mathbf{r}_{\mu}=(x,y,2R)$ and $\mathbf{r}=(x+100a, y, 2R)$, with the four different combinations of $x$ and $y$ indicated in the upper schematics. Panels (c) and (d) show, respectively, the spectrum for the cases with $\mathbf{r}_{\mu}=(0,0,z)$, $\mathbf{r}=(100a,0,z)$ and $\mathbf{r}_{\mu}=(0.5a,0.5a,z)$, $\mathbf{r}=(100.5a,0.5a,z)$, for three different values of $z$, as indicated by the legends. All of the results are normalized to the amplitude of the corresponding Green tensor of vacuum.}  \label{fig3}
\end{center}
\end{figure*}

The results displayed in Figure~\ref{fig2} demonstrate that $G^{\rm EE}_{zz}(\mathbf{r},\mathbf{r}_{\mu})$ is the dominant component of the Green tensor for the arrays under consideration. Therefore, in the following, we focus our analysis on this component. In Figure~\ref{fig3}(a), we plot the spectrum of $|G^{\rm EE}_{zz}(\mathbf{r},\mathbf{r}_{\mu})|/|G^{\rm EE}_{0,zz}(\mathbf{r}-\mathbf{r}_{\mu})|$ for $\mathbf{r}_{\mu}=(0,0,2R)$ and $\mathbf{r}=(\Delta x, 0, 2R)$ with $\Delta x=100a$ (red curve), $300a$ (black curve), and $600a$ (gray curve). As expected from the analysis above, once the lattice resonance begins to contribute (\textit{i.e.}, for $\lambda>832\,$nm), it produces a very large enhancement of the amplitude of the Green tensor of the array. The enhancement increases with $\Delta x$, reaching a peak of $\sim90$ times the value of the Green tensor of vacuum for $\Delta x=600a$. In Figure~\ref{fig3}(a), as in the rest of the figures of this article, we assume $a=800\,$nm and $R=100\,$nm. However, in Figure~S2 of the Appendix we analyze similar results for arrays with other values of $a$ and $R$. In all of the cases, for a given $a$, the spectral position of the maximum value of the amplitude of the Green tensor shifts to larger wavelengths as $R$ increases. This behavior is consistent with the redshift of the lattice resonance onset for increasing $R/a$ described in previous works \cite{ama68,ama72}. Furthermore, for each value of $a$, there is an optimum value of $R$ that produces the largest enhancement of the Green tensor amplitude.
 
Another important aspect to analyze is the dependence of the Green tensor of the array on the position of $\mathbf{r}_{\mu}$ and $\mathbf{r}$ within their respective unit cells. Figure~\ref{fig3}(b) shows the spectrum of $|G^{\rm EE}_{zz}(\mathbf{r},\mathbf{r}_{\mu})|/|G^{\rm EE}_{0,zz}(\mathbf{r}-\mathbf{r}_{\mu})|$ for $\mathbf{r}_{\mu}=(x,y,2R)$ and $\mathbf{r}=(x+100a, y, 2R)$ with the four different combinations of $x$ and $y$ depicted in the upper inset. The configuration in which both $\mathbf{r}_{\mu}$ and $\mathbf{r}$ are located above a nanoparticle (red curve) results in the largest values of the Green tensor, followed by the case in which both points lie at the center of the unit cell (green curve). The least favorable configurations correspond to $\mathbf{r}_{\mu}$ and $\mathbf{r}$ located in between two of the nanoparticles, either along the $x$-axis (blue curve) or the $y$-axis (yellow curve). These results can be explained as a combination of two different factors: On one hand, the excitation of the lattice resonance, as well as the field that it produces, become stronger as $\mathbf{r}_{\mu}$ and $\mathbf{r}$ get closer to the nanoparticles, thus resulting in a larger value of the Green tensor. On the other hand, configurations in which $\mathbf{r}_{\mu}$ and $\mathbf{r}$ are located in highly symmetrical points also favor a larger value of the Green tensor, since they minimize the cancelations due to phase differences in the excitation of the nanoparticles, as well as in the field that they produce. This last factor explains why the results for $(x,y)=(0.5a,0.5a)$ (green curve) are larger than those of $(x,y)=(0.5a,0)$ (blue curve) and $(x,y)=(0,0.5a)$ (yellow curve).

The value of the Green tensor also depends on the component of $\mathbf{r}_{\mu}$ and $\mathbf{r}$ along the direction perpendicular to the array (\textit{i.e.},  the $z$-axis). We explore this dependence in Figures~\ref{fig3}(c) and (d), where we plot the spectrum of $|G^{\rm EE}_{zz}(\mathbf{r},\mathbf{r}_{\mu})|/|G^{\rm EE}_{0,zz}(\mathbf{r}-\mathbf{r}_{\mu})|$ for $\mathbf{r}_{\mu}=(x,y,z)$ and $\mathbf{r}=(x+100a, y, z)$ with different values of $z$. Specifically, panels (c) and (d) show results for $(x,y)=(0,0)$ and $(x,y)=(0.5a,0.5a)$, respectively, with $z$ ranging from $4R$ (lighter curves) to $1.5R$ (darker curves). Analyzing these results, we observe that, in both cases, the amplitude of the Green tensor grows as $z$ decreases. However, while this growth accelerates for $(x,y)=(0,0)$ as $\mathbf{r}_{\mu}$ and $\mathbf{r}$ approach the nanoparticles, it saturates for $(x,y)=(0.5a,0.5a)$. 

Importantly, we have verified the accuracy of the dipole model by calculating the local density of states (LDOS) \cite{ama60} induced by a single metallic nanosphere, with the same characteristics as those of the arrays under consideration, and benchmarking it against full numerical solutions of Maxwell's equations. The results of this comparison are shown in Figure~S3 of the Appendix. As we discussed above, a fully numerical calculation of the Green tensor of the array using a FEM or FDTD solver of Maxwell's equations is extremely challenging due to the lack of periodicity of the problem. 

\begin{figure*}
\begin{center}
\includegraphics[width=150mm,angle=0]{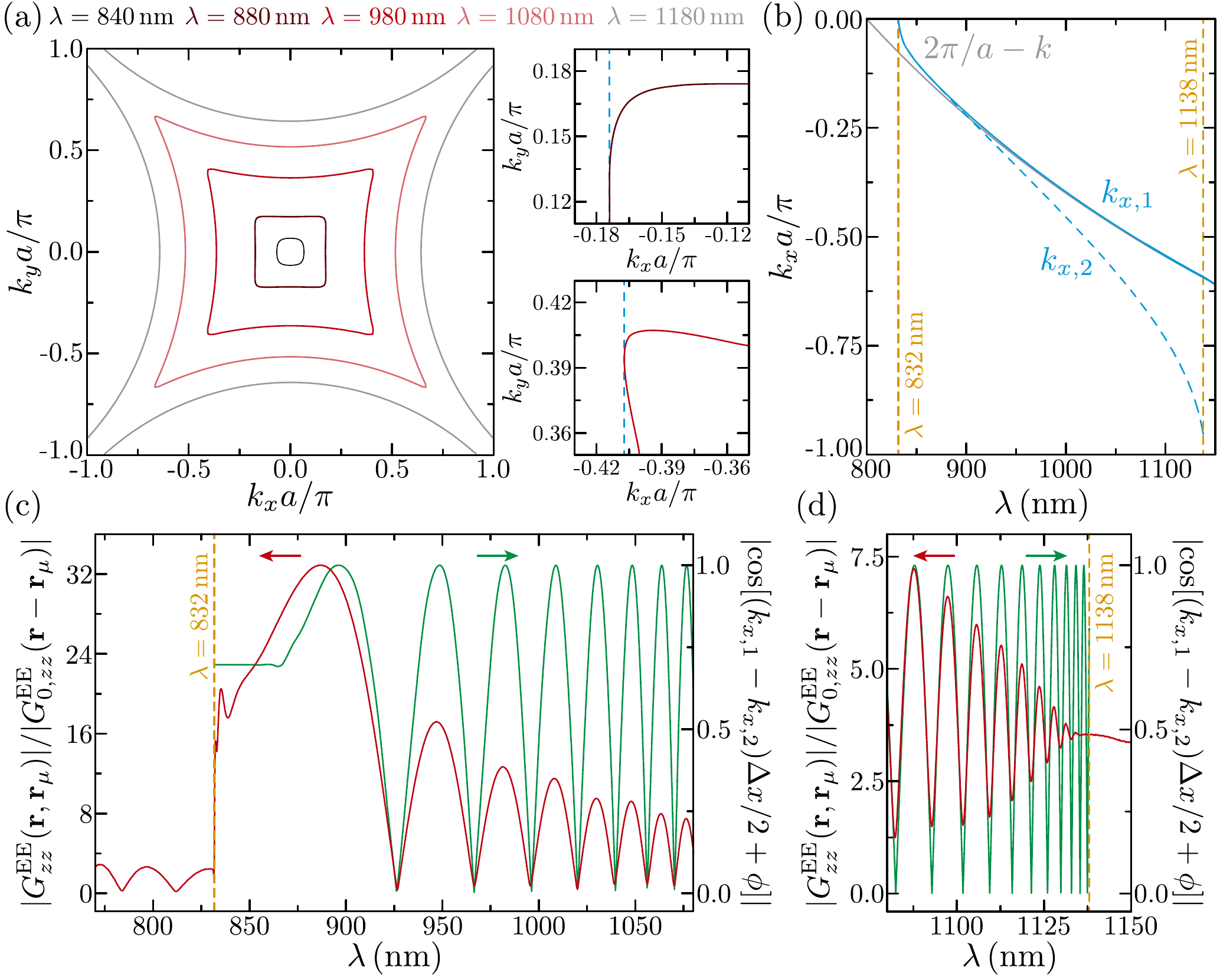}
\caption{Analysis of the spectral characteristics of the Green tensor of a periodic array. (a) Iso-contours showing the position of the lattice resonance peak within the first Brillouin zone for different wavelengths. The panels on the right display a zoom of the iso-contours for $\lambda=880\,$nm (upper panel) and $\lambda=980\,$nm (lower panel). The blue dashed lines indicate $k_{x,2}$. (b) Value of $k_{x,1}$ (blue solid curve) and $k_{x,2}$ (blue dashed curve) as a function of wavelength. The gray curve represents the value of $2\pi/a-k$. (c,d) Normalized amplitude of the $zz$ component of the Green tensor of  the array connecting $\mathbf{r}_{\mu}=(0,0,2R)$ and $\mathbf{r}=(100a,0,2R)$ (red curves, left scales) and results of the analytical model of Eq.~\ref{eqG3} (green curves, right scales). The yellow dashed lines in panels (b)-(d) mark the onset of the lattice resonance at the $\Gamma$ point and its cutoff at the M point, as shown in the insets of Figure~\ref{fig1}(c). In all panels, we assume $a=800\,$nm and $R=100\,$nm.} \label{fig4}
\end{center}
\end{figure*}

The spectra displayed in Figure~\ref{fig3} show that the start of the contribution of the lattice resonance to the Green tensor is not influenced by $\mathbf{r}_{\mu}$ and $\mathbf{r}$. However, the oscillations of its amplitude are strongly dependent on $\Delta x$, but are not affected by the position of $\mathbf{r}_{\mu}$ and $\mathbf{r}$ within their corresponding unit cells. In order to gain more insight into these behaviors, we start by invoking the following property of the lattice sum tensor: $\mathbfcal{G}(\mathbf{k}_{\parallel},\mathbf{R}_i+\boldsymbol{\rho})=\mathbfcal{G}(\mathbf{k}_{\parallel},\boldsymbol{\rho})e^{i \mathbf{k}_{\parallel}\cdot \mathbf{R}_i}$, where $\mathbf{R}_i$ is an arbitrary lattice vector. Using this relationship, we can rewrite Eq.~\ref{eqG} as
\begin{align}
\mathbf{G}(\mathbf{r},\mathbf{r}_{\mu})={}&\frac{a^2}{4\pi^2}  \int_{\rm 1BZ} d\mathbf{k}_{\parallel}  e^{i\mathbf{k}_{\parallel}\cdot (\mathbf{R}-\mathbf{R}_{\mu})} \nonumber \\ &\times \mathbfcal{G}(\mathbf{k}_{\parallel},\boldsymbol{\rho})  \mathbfcal{A}(\mathbf{k}_{\parallel})   \mathbfcal{G}(\mathbf{k}_{\parallel},-\boldsymbol{\rho}_{\mu}),\label{eqG2}
\end{align}
where we have defined $\boldsymbol{\rho}=\mathbf{r}-\mathbf{R}$ and $\boldsymbol{\rho}_{\mu}=\mathbf{r}_{\mu}-\mathbf{R}_{\mu}$, with $\mathbf{R}$ and $\mathbf{R}_{\mu}$ being the lattice vectors corresponding to the unit cells in which $\mathbf{r}_{\mu}$ and $\mathbf{r}$ are located. 
From their definition, it is clear that the in-plane components of $\boldsymbol{\rho}$ and $\boldsymbol{\rho}_{\mu}$ are located in the same unit cell.
Therefore, Eq.~\ref{eqG2} shows that the integral that defines the Green tensor of the array can be separated into two different factors: (i) an oscillating exponential that only depends on the separation between the unit cells in which $\mathbf{r}_{\mu}$ and $\mathbf{r}$ are located, and (ii) a term containing the response of the array, which only depends on the position of  $\mathbf{r}_{\mu}$ and $\mathbf{r}$ within the unit cell. 

As discussed in Figure~\ref{fig1}, the response of the array is determined by the characteristics of its lattice resonances. Figure~\ref{fig4}(a) shows the iso-contours indicating the position, within the first Brillouin zone, of the lowest-order lattice resonance of an array with $a=800\,$nm and $R=100\,$nm for the different values of the wavelength indicated by the labels. These iso-contours are obtained by finding the maximum value of $\mathbfcal{E}_{\mathbf{\hat{z}}}$ (see Eq.~\ref{eqExt}), and can be used to obtain an approximate expression for the Green tensor of the array.  
In particular, focusing again on the configuration with $\mathbf{r}_{\mu}=(0,0,2R)$ and $\mathbf{r}=(\Delta x,0,2R)$, and assuming that the response of the array is completely dominated by the lattice resonance, so that only the iso-contours contribute to the integral of Eq.~\ref{eqG2}, we can write
\begin{align}
G^{\rm EE}_{zz}(\mathbf{r},\mathbf{r}_{\mu}) \sim{}& e^{i k_{x,1} \Delta x} + e^{i k_{x,2} \Delta x}e^{ -i2\phi}  \nonumber \\ ={}& 2 e^{i (k_{x,1} + k_{x,2})\Delta x/2}e^{-i\phi} \nonumber \\ &\times {\cos}[(k_{x,1}-k_{x,2})\Delta x/2 + \phi]. \label{eqG3}
\end{align}
Here, $\phi$ is a constant phase, while $k_{x,1}$ and $k_{x,2}$ represent, in the spirit of the stationary phase approximation, the points of the lattice resonance iso-contours that have a zero derivative with respect to $k_y$. These are the only points that contribute to the integral, since the rapid oscillation of the exponential factor cancels the contribution of the rest of the iso-contour. The first of these points, $k_{1,x}$, corresponds to the value of $k_x$ at which the iso-contour intercepts the $k_{x}$-axis. The other one, $k_{x,2}$, as shown by the blue dashed lines in the right panels of Figure~\ref{fig4}(a), is located near the corners of the iso-contour. Figure~\ref{fig4}(b) shows the value of $k_{x,1}$ and $k_{x,2}$ as a function of wavelength. As expected from the shape of the iso-contours,  $k_{x,1}$ and $k_{x,2}$ take identical values for wavelengths near the onset of the lattice resonance at the $\Gamma$ point (\textit{i.e.}, $\lambda=832\,$nm). However, as $\lambda$ grows, their values become increasingly different. Importantly, the value of $k_{2,x}$ saturates as the wavelength approaches the cutoff of the lattice resonance at the M point (\textit{i.e.}, $\lambda = 1138\,$nm). These limits at the $\Gamma$ and M points are indicated with yellow dashed lines in Figure~\ref{fig4}(b), as well as in the insets of Figure~\ref{fig1}(c). It is important to mention that, in our analysis, we only consider the $k_{x,1}$ and $k_{x,2}$ located in the negative part of the $k_x$-axis. The reason is that the lowest order lattice resonance has a negative group velocity, as can be seen in Figure~\ref{fig1}(c), and therefore only the components with negative $k_{x}$ contribute to the Green tensor connecting the points $\mathbf{r}_{\mu}=(0,0,2R)$ and $\mathbf{r}=(\Delta x,0,2R)$ for positive values of $\Delta x$. This is confirmed numerically in Figure~S4 of the Appendix.

The analytical approximation given in Eq.~\ref{eqG3} predicts two different oscillatory behaviors. 
First, the exponential factor oscillates as the value of $(k_{1,x}+k_{2,x})\Delta x/2$ changes with wavelength. Since, as can be inferred from Figure~\ref{fig4}(b), the value of $(k_{1,x}+k_{2,x})/2$ is very similar to $2\pi/a-k$ (solid gray curve), especially for wavelengths below $\sim 1050\,$nm, this factor explains the fast oscillation of the real and imaginary parts of $\mathbf{G}^{\rm EE}(\mathbf{r},\mathbf{r}_{\mu})$ observed in Figure~\ref{fig2}. 
Second, the cosine factor produces a slower oscillation determined by the change of $(k_{1,x}-k_{2,x})\Delta x/2$ with $\lambda$.  As we analyze in Figure~\ref{fig4}(c), this oscillation reproduces that of the amplitude of the Green tensor of the array discussed in Figures~\ref{fig2} and \ref{fig3}. 
In particular, the green curve (right scale) displays the value of $|{\cos}[(k_{x,1}-k_{x,2})\Delta x/2 +\phi]|$, with  $k_{x,1}$ and $k_{x,2}$ taken from panel (b), $\Delta x = 100a$, and $\phi=0.8$, while the red curve (left scale) shows $|G^{\rm EE}_{zz}(\mathbf{r},\mathbf{r}_{\mu})|/|G^{\rm EE}_{0,zz}(\mathbf{r}-\mathbf{r}_{\mu})|$. Comparing these two curves, we observe that the cosine factor of Eq.~\ref{eqG3} perfectly matches the oscillations of the amplitude of the Green tensor above the onset of the lattice resonance (\textit{i.e.}, $\lambda = 832\,$nm). Furthermore, since $k_{2,x}$ saturates at $\lambda = 1138\,$nm, it is expected that these oscillations disappear beyond that cutoff. This prediction is  confirmed by the results plotted in Figure~\ref{fig4}(d), which extend those of panel (c) to the wavelength range around $1138\,$nm. Clearly, the oscillations of the amplitude of the Green tensor vanish as the wavelength approaches the cutoff indicated by the yellow dashed line, a behavior that is perfectly captured by the analytical approximation.

Then, from the results discussed in Figure~\ref{fig4} and the analytical approximation of Eq.~\ref{eqG3}, we conclude that the contribution of the lowest-order lattice resonance to the Green tensor of the array occurs through a combination of parallel wavevectors. One of them is always pointing along the direction connecting $\mathbf{r}$ with $\mathbf{r}_{\mu}$, while the value of the other varies with the wavelength. When the wavevectors are equal, as is the case for $\lambda \sim 880\,$nm, the amplitude of the Green tensor reaches its maximum value. However, as they become different, their contributions interfere, resulting in the oscillation of the amplitude of the Green tensor. Although these conclusions arise from the analysis of a particular set of $\mathbf{r}_{\mu}$ and $\mathbf{r}$, they can be readily extended to other configurations using symmetry arguments.

\begin{figure}
\begin{center}
\includegraphics[width=75mm,angle=0]{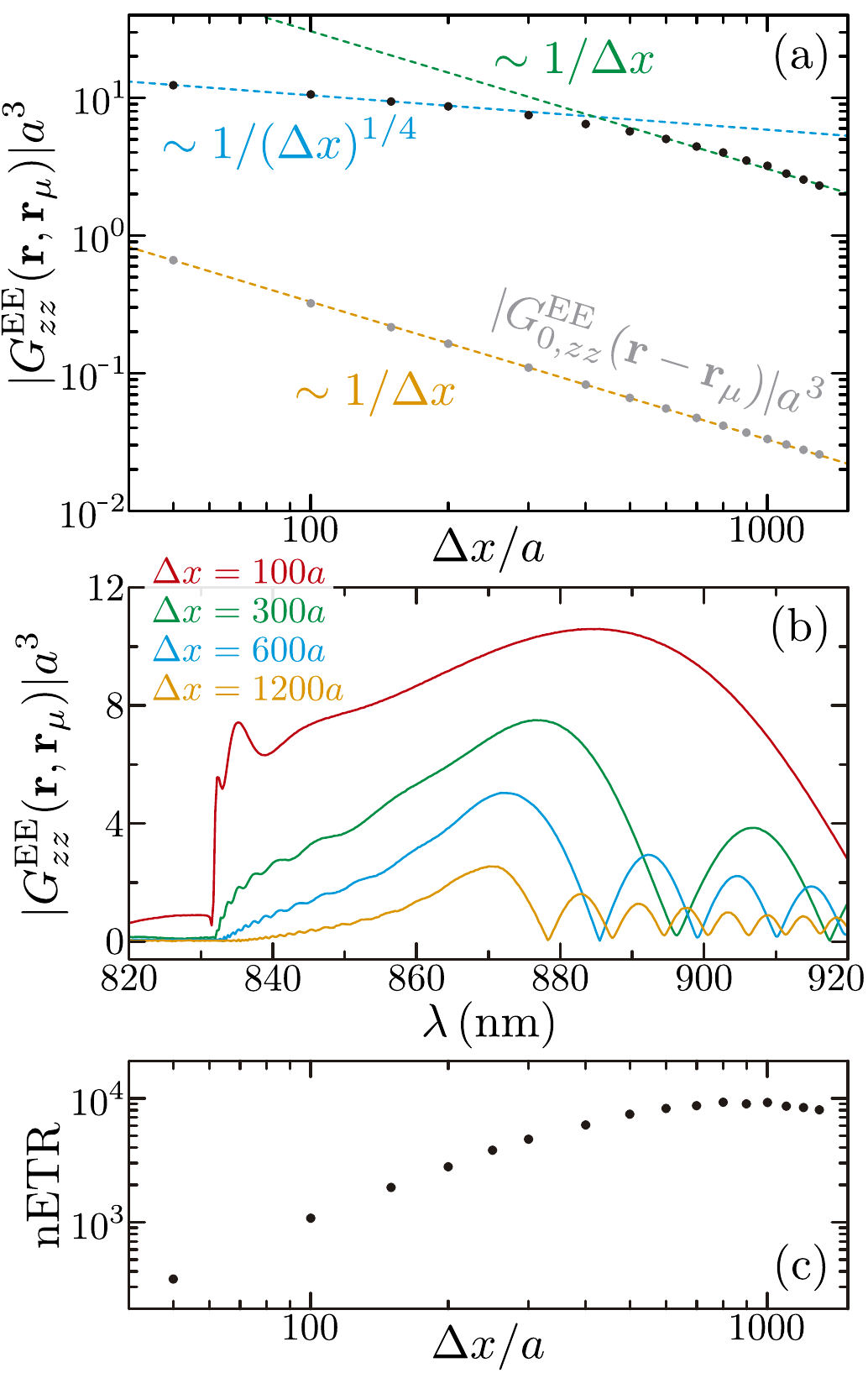}
\caption{Analysis of the dependence with $|\mathbf{r}-\mathbf{r}_{\mu}|$ of the Green tensor of a periodic array with $a=800\,$nm and $R=100\,$nm. The black dots in panel (a) show $|G^{\rm EE}_{zz}(\mathbf{r},\mathbf{r}_{\mu})|$, for $\mathbf{r}_{\mu}=(0,0,2R)$ and $\mathbf{r}=(\Delta x,0,2R)$ as a function of $\Delta x$. For each $\Delta x$, we calculate $|G^{\rm EE}_{zz}(\mathbf{r},\mathbf{r}_{\mu})|$ at the wavelength that produces the largest value. For comparison, the gray dots represent $|G^{\rm EE}_{0,zz}(\mathbf{r}-\mathbf{r}_{\mu})|$, evaluated at the same positions and wavelengths. The dashed lines mark different scaling functions as indicated by the corresponding labels. (b) Spectrum of $|G^{\rm EE}_{zz}(\mathbf{r},\mathbf{r}_{\mu})|$ for the different values of $\Delta x$ shown in the legend. (c) Normalized energy transfer rate calculated from the data shown in panel (a) as ${\rm nETR}=|G^{\rm EE}_{zz}(\mathbf{r},\mathbf{r}_{\mu})|^2/|G^{\rm EE}_{0,zz}(\mathbf{r}-\mathbf{r}_{\mu})|^2$. } \label{fig5}
\end{center}
\end{figure}

In order to complete our characterization of the Green tensor of the array, we analyze, in Figure~\ref{fig5}(a), its dependence with the distance between $\mathbf{r}_{\mu}$ and $\mathbf{r}$. Specifically, we consider an array with $a=800\,$nm and $R=100\,$nm, and take $\mathbf{r}_{\mu}=(0,0,2R)$ and $\mathbf{r}=(\Delta x,0,2R)$. We use black dots to plot the value of $|G^{\rm EE}_{zz}(\mathbf{r},\mathbf{r}_{\mu})|$ as a function of $\Delta x$. For each value of $\Delta x$, we evaluate $|G^{\rm EE}_{zz}(\mathbf{r},\mathbf{r}_{\mu})|$ at the wavelength for which it reaches its maximum value. As a reference, we also plot, using gray dots, the value of the amplitude of the Green tensor of vacuum $|G^{\rm EE}_{0,zz}(\mathbf{r}-\mathbf{r}_{\mu})|$, evaluated at the same positions and wavelengths. Analyzing these results, we observe that, while the Green tensor of vacuum always decays as $(\Delta x)^{-1}$, as expected from the far-field character of  the distances under consideration, $|G^{\rm EE}_{zz}(\mathbf{r},\mathbf{r}_{\mu})|$ displays two distinct behaviors. For $\Delta x \lesssim 400a$, it decays at a much slower rate, following an approximate dependence of $\sim(\Delta x)^{-1/4}$. However, for $\Delta x \gtrsim 400a$, the decay accelerates to $\sim (\Delta x)^{-1}$, similar to that of the Green tensor of vacuum. This change of behavior can be understood by looking at the spectra of $|G^{\rm EE}_{zz}(\mathbf{r},\mathbf{r}_{\mu})|$ plotted in Figure~\ref{fig5}(b). 
There we see how, as $\Delta x$ increases, the first minimum caused by the oscillations of the Green tensor shifts towards smaller wavelengths. 
For $\Delta x \gtrsim 400a$, this minimum reaches the wavelengths at which the amplitude of the Green tensor achieves its maximum value ($\lambda \sim 880\,$nm) and forces it to decrease, thus producing the faster decay rate observed in panel (a).

The results shown in panel (a) demonstrate that the contribution of the lattice resonance produces a large enhancement of the Green tensor of the array as compared with its vacuum counterpart. This enhancement can mediate the transfer of energy between dipole emitters placed in the vicinity of the array. In order to quantify this effect, we plot, in Figure~\ref{fig5}(c), the normalized energy transfer rate, defined as \cite{NH06,MMG10}
\begin{equation}
{\rm nETR} = \frac{|G^{\rm EE}_{zz}(\mathbf{r},\mathbf{r}_{\mu})|^2}{|G^{\rm EE}_{0,zz}(\mathbf{r}-\mathbf{r}_{\mu})|^2}. \nonumber
\end{equation}
This quantity measures the enhancement of the energy transfer between two electric dipole emitters located at $\mathbf{r}_{\mu}$ and $\mathbf{r}$ provided by the array. We assume that both dipoles are oriented along the $z$-axis since, as discussed before, that is the optimum configuration to maximize the contribution of the array. The results shown in Figure~\ref{fig5}(c) clearly demonstrate that, thanks to the contribution of the lattice resonance, the array under investigation produces a nETR with values in the range of $10^3$ to $10^4$ for distances of hundreds to thousands of periods. These results confirm that periodic arrays of metallic nanostructures are capable of enhancing the long-range coupling between dipole emitters.

\section{Conclusions}

In summary,  we have performed a detailed investigation of the coupling between dipole emitters mediated by the lattice resonances of a periodic array of metallic nanoparticles. To do so,  we have derived a closed expression for the Green tensor of the array using a rigorous coupled dipole model and used it to analyze its spectral and spatial characteristics. We have focused on the electric-electric term of the Green tensor, which represents the electric field produced by the array when excited by a unit electric dipole, and analyzed both its $yy$ and the $zz$ components. By doing so, we have found that the latter reaches much larger values due to the contribution of the out-of-plane lattice resonance, which displays much lower radiative losses than its in-plane counterpart.  
Through the analysis of the spectrum of the Green tensor connecting different pairs of points $\mathbf{r}_{\mu}$ and $\mathbf{r}$, we have found that, in addition to a fast oscillation of its real and imaginary parts arising from its far-field nature, the amplitude of the Green tensor displays a slower oscillation. This oscillation depends on $|\mathbf{r}-\mathbf{r}_{\mu}|$, but not on the position of these points within the unit cells in which they are located. We have explained this behavior as the result of an interference process produced by the excitation of a lattice resonance with different parallel wavevectors.  By comparing the Green tensor of the array with its vacuum counterpart, we have found that the contribution of the lattice resonance results in extraordinarily large values that decay with $|\mathbf{r}-\mathbf{r}_{\mu}|$ at a much slower rate. This demonstrates that the lattice resonances of periodic arrays of metallic nanoparticles can mediate an efficient long-range coupling between dipole emitters placed in their vicinity. Although, in this article, we have focused on arrays of metallic nanoparticles, our theoretical approach is also valid for investigating other systems, such as arrays of dielectric nanostructures \cite{AS21}. Furthermore, our analysis can be readily applied to periodic arrays of atoms by using the appropriate polarizability \cite{BGA16,SWL17,GGV19,AGA20,RWR20}. The results of this article expand the fundamental knowledge of lattice resonances and pave the way for the use of periodic arrays of metallic nanostructures as platforms to enhance the long-range coupling between dipole emitters.

\acknowledgments
This work has been sponsored by the Spanish Ministry for Science and Innovation (Grant No. TEM-FLU PID2019-109502GA-I00) and the U.S. National Science Foundation (Grant No. DMR-1941680). L.Z. acknowledges support from the Department of Energy Computational Science Graduate Fellowship (Grant No. DE-SC0020347). We would also like  to thank the UNM Center for Advanced Research Computing, supported in part by the U.S. National Science Foundation, for providing some of the computational resources used in this work.  A.C-G. and A.I.F-D. acknowledge funding from Spanish MINECO under contract No. MDM-2014-0377-16-4 and from the Spanish Ministry for Science and Innovation under contract RTI2018-099737-B-I00 and through the ``Mar\'{i}a de Maeztu'' programme for Units of Excellence in R\&D (CEX2018-000805-M).

\onecolumngrid
\appendix
\section{Appendix}\label{ap}
\renewcommand{\thefigure}{S\arabic{figure}}
\setcounter{figure}{0} 

\subsection{Derivation of the Polarizability of the Array}
As stated in the main text, we use a coupled dipole model \cite{ZKS03,paper090,TD12,ama59,KK19} and describe each of the nanoparticles of the array as a point dipole with both electric $\mathbf{p}$ and magnetic $\mathbf{m}$ components. The dipole induced in the nanoparticle of the array located at $\mathbf{R}_i$, when excited by an external electromagnetic field with electric and magnetic amplitudes $\mathbf{E}_i$ and $\mathbf{B}_i$, respectively, can be written as
\begin{equation}
\left[\begin{matrix} \mathbf{p}_i \\ \mathbf{m}_i \end{matrix}\right] =  \left[\begin{matrix}\boldsymbol{\alpha}^{\rm E} & 0\\ 0 &\boldsymbol{\alpha}^{\rm M}\end{matrix}\right] \left( \left[\begin{matrix}\mathbf{E}_i\\ \mathbf{B}_i\end{matrix}\right]+\sum_{j \neq i} \left[ \begin{matrix} \mathbf{G}_0^{\rm EE}(\mathbf{R}_i-\mathbf{R}_j) & \mathbf{G}_0^{\rm EM}(\mathbf{R}_i-\mathbf{R}_j) \\ \mathbf{G}_0^{\rm ME}(\mathbf{R}_i-\mathbf{R}_j) & \mathbf{G}_0^{\rm MM}(\mathbf{R}_i-\mathbf{R}_j) \end{matrix}\right] \left[\begin{matrix}\mathbf{p}_j\\ \mathbf{m}_j\end{matrix}\right]\right).\nonumber
\end{equation}
Here, $\boldsymbol{\alpha}^{\rm E}$ and $\boldsymbol{\alpha}^{\rm M}$ are the electric and magnetic polarizability tensors of the nanoparticles, while $\mathbf{G}_0^{\rm EE}(\mathbf{r})=\mathbf{G}_0^{\rm MM}(\mathbf{r})=[k^2\mathcal{I}_{3\times3}+\nabla\nabla] e^{i k |\mathbf{r}|}/|\mathbf{r}|$ and $\mathbf{G}_0^{\rm EM}(\mathbf{r})=-\mathbf{G}_0^{\rm ME}(\mathbf{r})=ik\nabla\times e^{i k |\mathbf{r}|}/|\mathbf{r}|$ represent the different components of the Green tensor of vacuum, with $k=2\pi /\lambda$ being the wavenumber of light (notice that we use Gaussian units).
Taking advantage of the periodicity of the array and using the Fourier transform defined as $v_i = \frac{a^2}{4\pi^2}\int_{\rm 1BZ} d\mathbf{k}_{\parallel} v(\mathbf{k}_{\parallel})e^{i \mathbf{k}_{\parallel}\cdot \mathbf{R}_{i}}$, where $a$ is the array period and $\rm 1BZ$ stands for the first Brillouin zone, we can write the following self-consistent equation for the $\mathbf{k}_{\parallel}$ components of the dipole induced in the nanoparticles
 \begin{equation}
 \left[\begin{matrix}\mathbf{p}(\mathbf{k}_{\parallel})\\ \mathbf{m}(\mathbf{k}_{\parallel})\end{matrix}\right] =  \left[\begin{matrix}\boldsymbol{\alpha}^{\rm E} & 0\\ 0 &\boldsymbol{\alpha}^{\rm M}\end{matrix}\right] \left(\left[\begin{matrix}\mathbf{E}(\mathbf{k}_{\parallel})\\ \mathbf{B}(\mathbf{k}_{\parallel})\end{matrix}\right]+\left[ \begin{matrix} \mathbfcal{G}^{\rm EE}(\mathbf{k}_{\parallel},0) & \mathbfcal{G}^{\rm EM}(\mathbf{k}_{\parallel},0) \\ -\mathbfcal{G}^{\rm EM}(\mathbf{k}_{\parallel},0) & \mathbfcal{G}^{\rm EE}(\mathbf{k}_{\parallel},0) \end{matrix}\right] \left[\begin{matrix}\mathbf{p}(\mathbf{k}_{\parallel})\\ \mathbf{m}(\mathbf{k}_{\parallel})\end{matrix}\right]\right).\label{eqpk}
\end{equation}
In this expression,  $\mathbfcal{G}^{\nu}(\mathbf{k}_{\parallel},\mathbf{r})= \sideset{}{'} \sum_{i} \mathbf{G}_0^{\nu}(\mathbf{R}_i+\mathbf{r})e^{-i\mathbf{k}_{\parallel}\cdot \mathbf{R}_i}$ with $\nu = {\rm EE, EM}$ and the prime in the summation indicates that, if a term satisfies $\mathbf{R}_i+\mathbf{r}=0$, it is to be excluded. Equation~\ref{eqpk} can be solved as
\begin{equation}
\left[\begin{matrix}\mathbf{p}(\mathbf{k}_{\parallel})\\ \mathbf{m}(\mathbf{k}_{\parallel})\end{matrix}\right] =\mathbfcal{A}(\mathbf{k}_{\parallel}) \left[\begin{matrix}\mathbf{E}(\mathbf{k}_{\parallel})\\ \mathbf{B}(\mathbf{k}_{\parallel})\end{matrix}\right],\label{eqA}
\end{equation}
where 
\begin{equation}
\mathbfcal{A}(\mathbf{k}_{\parallel})=  \left(\left[\begin{matrix}\boldsymbol{\alpha}^{\rm E} & 0\\ 0 &\boldsymbol{\alpha}^{\rm M}\end{matrix}\right]^{-1} -\left[ \begin{matrix} \mathbfcal{G}^{\rm EE}(\mathbf{k}_{\parallel},0) & \mathbfcal{G}^{\rm EM}(\mathbf{k}_{\parallel},0) \\ -\mathbfcal{G}^{\rm EM}(\mathbf{k}_{\parallel},0) & \mathbfcal{G}^{\rm EE}(\mathbf{k}_{\parallel},0) \end{matrix}\right] \right)^{-1}, \label{eqA2}
\end{equation}
is the polarizability of the array.

\subsection{ Derivation of the $\mathbf{k}_{\parallel}$ Components of the Electromagnetic Field of a Point Dipole} 

Given a unit dipole with electric $\boldsymbol{\hat{\mu}}^{\rm E}$ and magnetic $\boldsymbol{\hat{\mu}}^{\rm M}$ components, oscillating at frequency $\omega$ and located at $\mathbf{r}_{\mu}$, we can use the array scanning method \cite{CJW07,LK16} to write its associated current as
\begin{equation}
\left[ \begin{matrix} \mathbf{j}^{\rm E}(\mathbf{r}) \\ \mathbf{j}^{\rm M}(\mathbf{r}) \end{matrix}\right] = \frac{a^2}{4\pi^2}\int_{\rm 1BZ} d\mathbf{k}_{\parallel} (-i \omega) \sum_{i} \delta(\mathbf{r}-\mathbf{r}_{\mu}+\mathbf{R}_i) e^{-i \mathbf{k}_{\parallel}\cdot\mathbf{R}_i} \left[ \begin{matrix} \boldsymbol{\hat{\mu}}^{\rm E} \\ \boldsymbol{\hat{\mu}}^{\rm M} \end{matrix}\right]. \label{eqjk}
\end{equation}
Then, taking into account that the electromagnetic field produced by the dipole current at the position of one of the nanoparticles of the array can be written as
\begin{equation}
\left[\begin{matrix}\mathbf{E}_i\\ \mathbf{B}_i\end{matrix}\right] = \frac{i}{\omega} \int d\mathbf{r}'   \left[ \begin{matrix} \mathbf{G}_0^{\rm EE}(\mathbf{R}_i-\mathbf{r}') & \mathbf{G}_0^{\rm EM}(\mathbf{R}_i-\mathbf{r}') \\ -\mathbf{G}_0^{\rm EM}(\mathbf{R}_i-\mathbf{r}') & \mathbf{G}_0^{\rm EE}(\mathbf{R}_i-\mathbf{r}') \end{matrix}\right]  \left[\begin{matrix}\mathbf{j}^{\rm E}(\mathbf{r}')\\ \mathbf{j}^{\rm M}(\mathbf{r}')\end{matrix}\right],\nonumber
\end{equation}
and using Eq.~\ref{eqjk}, we obtain
\begin{equation}
\left[\begin{matrix}\mathbf{E}(\mathbf{k}_{\parallel})\\ \mathbf{B}(\mathbf{k}_{\parallel})\end{matrix}\right] = \left[ \begin{matrix} \mathbfcal{G}^{\rm EE}(\mathbf{k}_{\parallel},-\mathbf{r}_{\mu}) & \mathbfcal{G}^{\rm EM}(\mathbf{k}_{\parallel},-\mathbf{r}_{\mu}) \\ -\mathbfcal{G}^{\rm EM}(\mathbf{k}_{\parallel},-\mathbf{r}_{\mu}) & \mathbfcal{G}^{\rm EE}(\mathbf{k}_{\parallel},-\mathbf{r}_{\mu}) \end{matrix}\right] \left[\begin{matrix} \boldsymbol{\hat{\mu}}^{\rm E} \\ \boldsymbol{\hat{\mu}}^{\rm M}\end{matrix}\right].\label{eqEk}
\end{equation}

\subsection{ Derivation of the Electromagnetic Field Produced by a Periodic Array of Dipoles}  

We can write the electromagnetic field produced at a point $\mathbf{r}$ outside of the array by a periodic array of dipoles as 
\begin{equation}
\left[\begin{matrix}\mathbf{E}(\mathbf{r})\\ \mathbf{B}(\mathbf{r})\end{matrix}\right] = \sum_{i} \left[ \begin{matrix} \mathbf{G}_0^{\rm EE}(\mathbf{r}-\mathbf{R}_i) & \mathbf{G}_0^{\rm EM}(\mathbf{r}-\mathbf{R}_i) \\ -\mathbf{G}_0^{\rm EM}(\mathbf{r}-\mathbf{R}_i) & \mathbf{G}_0^{\rm EE}(\mathbf{r}-\mathbf{R}_i) \end{matrix}\right]  \left[\begin{matrix}\mathbf{p}_i\\ \mathbf{m}_i\end{matrix}\right].\nonumber
\end{equation}
Then, expressing the dipoles in terms of their $\mathbf{k}_{\parallel}$ components, we have
\begin{equation}
\left[\begin{matrix}\mathbf{E}(\mathbf{r})\\ \mathbf{B}(\mathbf{r})\end{matrix}\right] = \frac{a^2}{4\pi^2}\int_{\rm 1BZ} d\mathbf{k}_{\parallel} \left[ \begin{matrix} \mathbfcal{G}^{\rm EE}(\mathbf{k}_{\parallel},\mathbf{r}) & \mathbfcal{G}^{\rm EM}(\mathbf{k}_{\parallel},\mathbf{r}) \\ -\mathbfcal{G}^{\rm EM}(\mathbf{k}_{\parallel},\mathbf{r}) & \mathbfcal{G}^{\rm EE}(\mathbf{k}_{\parallel},\mathbf{r}) \end{matrix}\right]  \left[\begin{matrix}\mathbf{p}(\mathbf{k}_{\parallel})\\ \mathbf{m}(\mathbf{k}_{\parallel})\end{matrix}\right].\nonumber
\end{equation}
Finally, substituting Eqs.~\ref{eqA} and \ref{eqEk} into the expression above, we obtain Eq.~\ref{eqE}.

\newpage

\begin{figure}
\begin{center}
\includegraphics[width=150mm,angle=0]{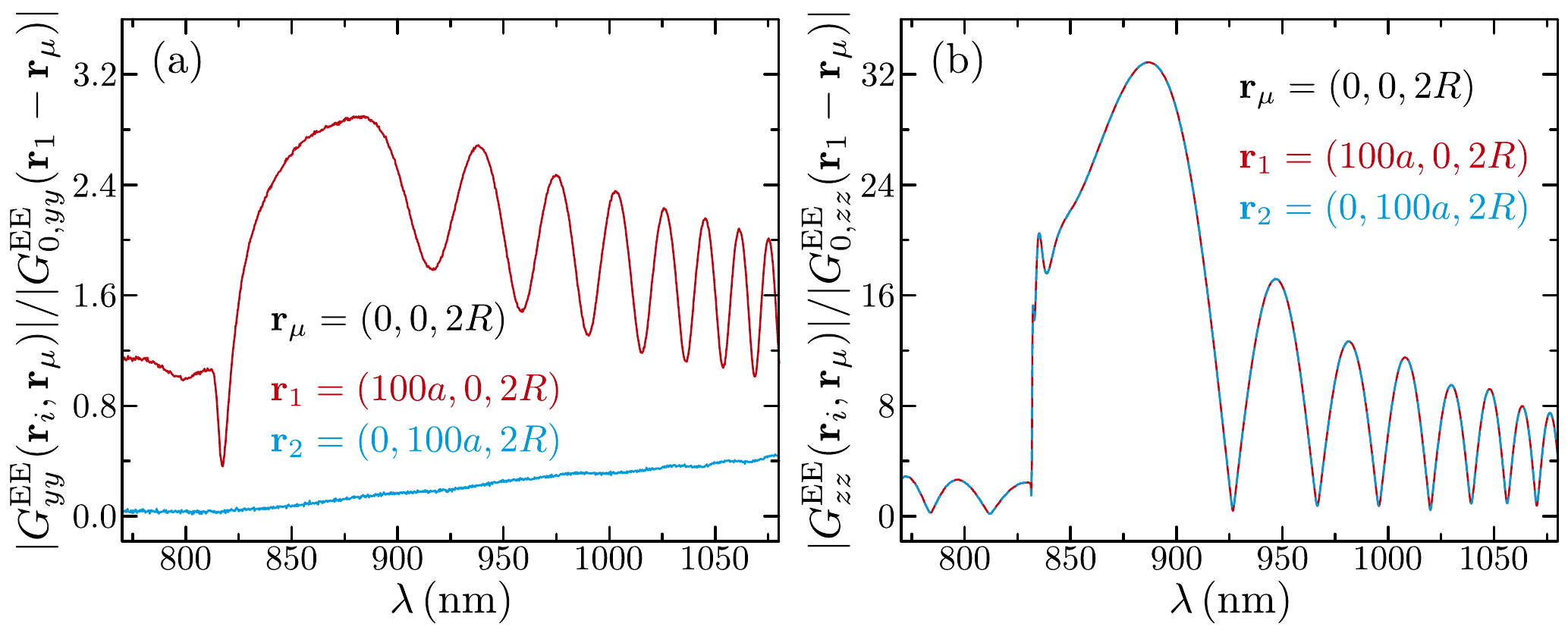}
\caption{Amplitude of the Green tensor of a periodic array with $a=800\,$nm and $R=100\,$nm connecting $\mathbf{r}_{\mu}=(0,0,2R)$ with either $\mathbf{r}_1=(100a,0,2R)$ (red curves) or $\mathbf{r}_2=(0,100a,2R)$ (blue curves). Panels (a) and (b) show the spectrum for the $yy$ and $zz$ components of the Green tensor normalized to $|G^{\rm EE}_{0,yy}(\mathbf{r}_{1}-\mathbf{r}_{\mu})|$ and $|G^{\rm EE}_{0,zz}(\mathbf{r}_{1}-\mathbf{r}_{\mu})|$, respectively. } \label{figS1}
\end{center}
\end{figure}

\begin{figure}
\begin{center}
\includegraphics[width=150mm,angle=0]{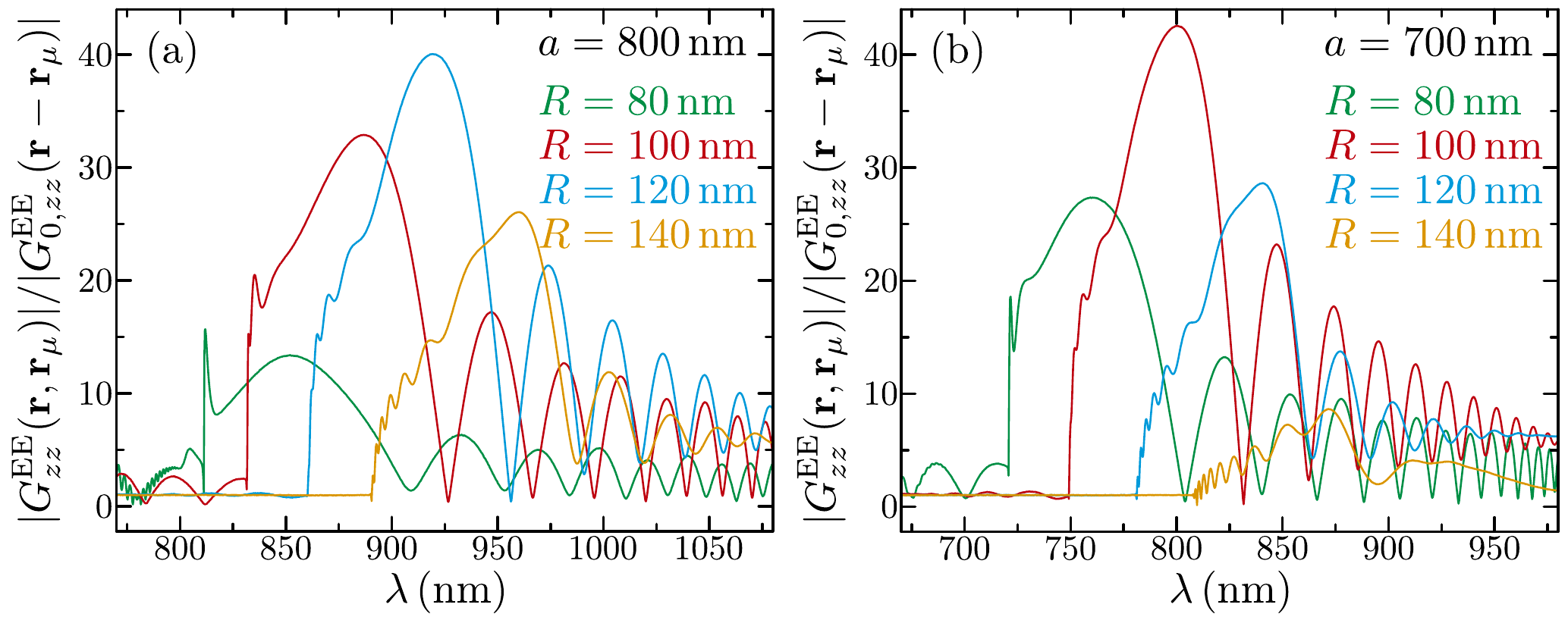}
\caption{Amplitude of the $zz$ component of the Green tensor of a periodic array connecting $\mathbf{r}_{\mu}=(0,0,2R)$ and $\mathbf{r}=(100a,0,2R)$ for different values of the array period $a$ and the nanoparticle radius $R$. Specifically, panels (a) and (b) show the spectrum for $a=800\,$nm and $a=700\,$nm, respectively, with $R$ ranging, in both cases, from $80$ to $140\,$nm.  All of the results are normalized to the amplitude of the corresponding Green tensor of vacuum. } \label{figS2}
\end{center}
\end{figure}

\begin{figure}
\begin{center}
\includegraphics[width=150mm,angle=0]{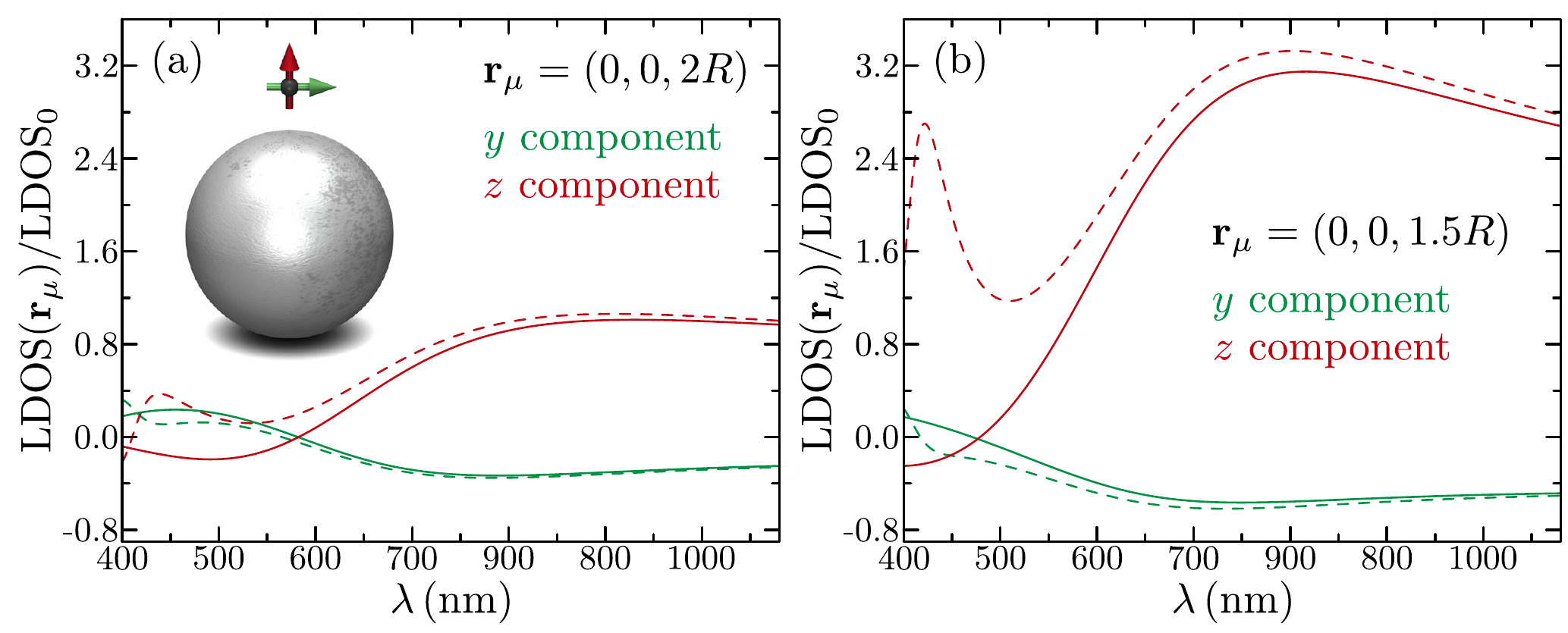}
\caption{Local density of photonic states (LDOS) induced by a metallic nanosphere of radius $R=100\,$nm at a point $\mathbf{r}_{\mu}$ normalized to the LDOS of vacuum LDOS$_0=4/(3c\lambda^2)$. Panels (a) and (b) show the LDOS spectrum at $\mathbf{r}_{\mu}=(0,0,2R)$ and $\mathbf{r}_{\mu}=(0,0,1.5R)$, respectively, with the origin of coordinates being set at the center of the nanoparticle. In both panels, the green (red) curves correspond to the $y$ component ($z$ component) of the LDOS.  Furthermore, dashed curves represent the results obtained from a full numerical solution of Maxwell's equations, while solid curves correspond to the results of the dipole model. Examining these spectra, we observe that the dipole model is in excellent agreement with the full numerical calculations for wavelengths above $\gtrsim 600\,$nm. As expected, the agreement deteriorates for smaller wavelengths due to the contribution of the quadrupole mode of the nanoparticle, which is not captured by the dipole model.} \label{figS3}
\end{center}
\end{figure}

\begin{figure}
\begin{center}
\includegraphics[width=150mm,angle=0]{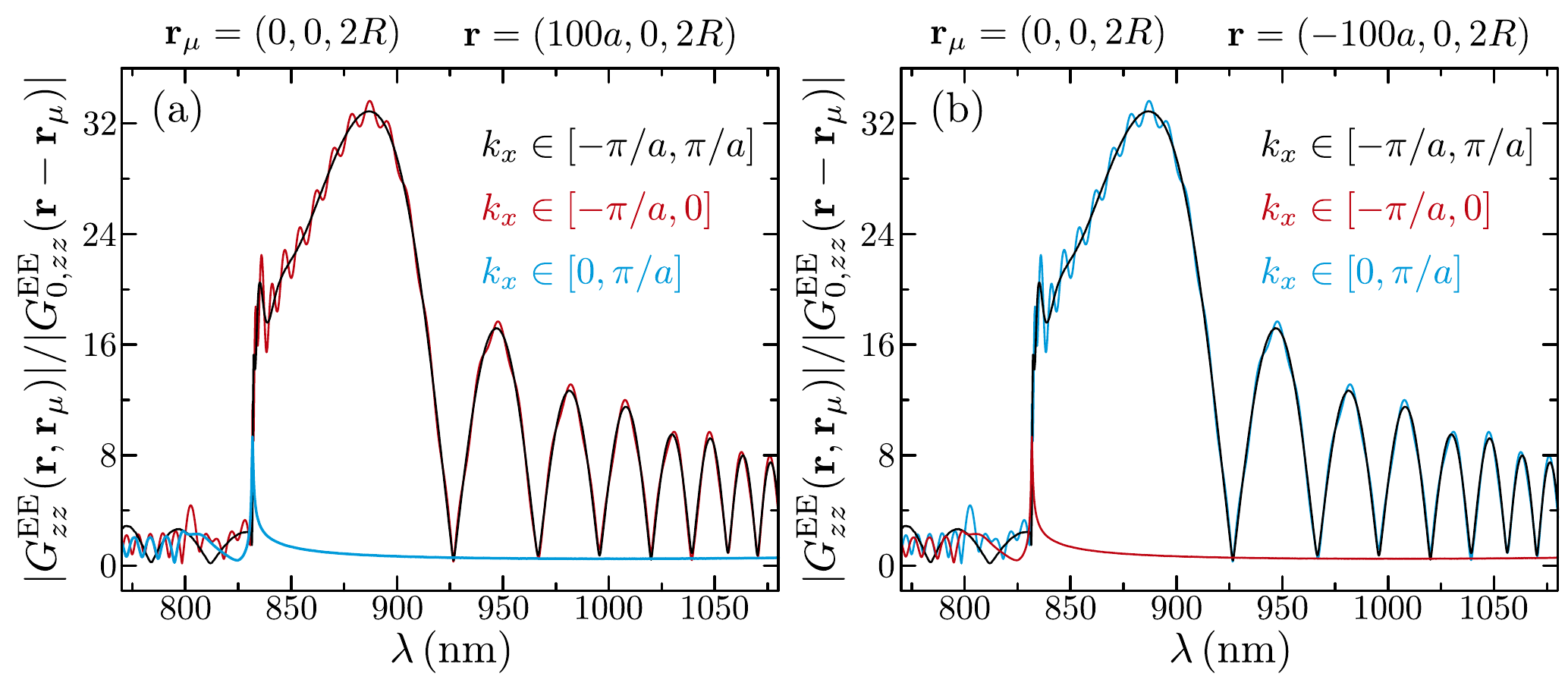}
\caption{Amplitude of the $zz$ component of the Green tensor of a periodic array with $a=800\,$nm and $R=100\,$nm connecting  $\mathbf{r}_{\mu}=(0,0,2R)$ with either $\mathbf{r}=(100a,0,2R)$ (a) or  $\mathbf{r}=(-100a,0,2R)$ (b). In both panels, the black curves represent the value of $|G^{\rm EE}_{zz}(\mathbf{r},\mathbf{r}_{\mu})|$ obtained from performing the integral of Eq.~3 over the entire first Brillouin zone, while the colored curves represent the results obtained by restricting the integral over $k_x$ to either $[-\pi/a,0]$ (red curves) or $[0,\pi/a]$ (blue curves).  All of the results are normalized to the amplitude of the corresponding Green tensor of vacuum. } \label{figS4}
\end{center}
\end{figure}


\end{document}